\documentclass{emulateapj}
\newcommand{\ifpp}[1]{#1}
\newcommand{\ifms}[1]{}

\newcommand{\unit}[1]{\ifmmode {\rm\ #1} \else {$\rm #1$} \fi}

\newcommand{\AlII}{{\rm Al{\sc ii}}}

\newcommand{\CaF}{\unit{CaF_2}}
\newcommand{\CIII}{{\rm C{\sc iii}}}
\newcommand{\CI}{{\rm C{\sc i}}}

\newcommand{\CIV}{{\rm C{\sc iv}}}

\newcommand{\COBE}{{\em COBE}}

\newcommand{\Einstein}{{\em Einstein}}

\newcommand{\FUSE}{{\em FUSE}}
\newcommand{\FeII}{{\rm Fe{\sc ii}}}

\newcommand{\HI}{{\rm H{\sc i}}}

\newcommand{\Htwo}{{\rm \unit{H_2}}}
\newcommand{\HeII}{{\rm He{\sc ii}}}

\newcommand{\IRAS}{{\em IRAS}}

\newcommand{\Lya}{\unit{Ly\,\alpha}}
\newcommand{\Lyb}{\unit{Ly\,\beta}}

\newcommand{\NI}{{\rm N{\sc i}}}
\newcommand{\NII}{{\rm N{\sc ii}}}
\newcommand{\NIII}{{\rm N{\sc iii}}}
\newcommand{\NIV}{{\rm N{\sc iv}}}

\newcommand{\NeVI}{{\rm Ne{\sc vi}}}
\newcommand{\Ntwo}{\unit{\rm N_2}}
\newcommand{\Ntwoplus}{\unit{\rm N_2^+}}

\newcommand{\OIII}{{\rm O{\sc iii}}}
\newcommand{\OIIIb}{{\rm O{\sc iii]}}}

\newcommand{\OIVb}{{\rm O{\sc iv]}}}
\newcommand{\OI}{{\rm O{\sc i}}}
\newcommand{\OVI}{{\rm O{\sc vi}}}

\newcommand{\ROSAT}{{\em ROSAT}}

\newcommand{\SPEAR}{{\em SPEAR}}
\newcommand{\SPEARFIMS}{{\em SPEAR/FIMS}}

\newcommand{\SiII}{{\rm Si{\sc ii}}}
\newcommand{\SiIV}{{\rm Si{\sc iv}}}

\newcommand{\angstrom}{\unit{\AA}}

\newcommand{\lu}{\unit{ph\ s^{-1}}\-\unit{cm^{-2}}\- \unit{sr^{-1}}}
\newcommand{\cu}{\unit{ph\ s^{-1}}\-\unit{cm^{-2}}\- \unit{sr^{-1}\- \angstrom^{-1}}}
\newcommand{\degrees}{\ifmmode {^{\circ}}\else {$^{\circ}$}\fi}
\newcommand{\degree}{\ifmmode {^{\circ}}\else {$^{\circ}$}\fi}

\newcommand{\doublet}{\ifmmode {\lambda\lambda} \else {$\lambda\lambda$} \fi}
\newcommand{\eV}{\unit{eV}}
\newcommand{\emunits}{\unit{{\rm cm}^{-6} {\rm pc}}}

\newcommand{\etal}{{et~al.~}}
\newcommand{\gt}{\unit{>}}
\newcommand{\gtsim}{\unit{_{\sim}^{>}}}

\newcommand{\simgt}{\unit{^>_\sim}}
\newcommand{\simlt}{\unit{^<_\sim}}
\newcommand{\lt}{\unit{<}}

\newcommand{\persqrcm}{\unit{cm^{-2}}}

\newcommand{\quarter}{\ifmmode {\frac{1}{4}} \else {$\frac{1}{4}$} \fi}

\newcommand{\singlet}{\ifmmode {\lambda} \else {$\lambda$} \fi}

\newcommand{\taudust}{\unit{\tau_{\rm dust}}}
\newcommand{\tentothe}[1]{\ifmmode {10^{#1}} \else {$10^{#1}$} \fi}
\newcommand{\tten}[1]{\ifmmode {\times 10^{#1}} \else {$\times 10^{#1}$} \fi}

\shorttitle{\SPEAR\ observations of the NEP}
\shortauthors{Korpela \etal}
\begin{document}
\title{Far-ultraviolet Observations of the North Ecliptic Pole with
\SPEAR}
\author{Eric J.~Korpela\altaffilmark{1}, Jerry Edelstein\altaffilmark{1},
Julia Kregenow\altaffilmark{1}, Kaori Nishikida\altaffilmark{1},
Kyoung-Wook Min\altaffilmark{2}, Dae-Hee Lee\altaffilmark{2,3},  Kwangsun
Ryu\altaffilmark{2}, Wonyong Han\altaffilmark{3}, Uk-Won Nam\altaffilmark{3},
Jang-Hyun Park\altaffilmark{3}}
\slugcomment{Accepted for publication in ApJ Letters on 19 December 2005}

\altaffiltext{1}{Space Sciences Laboratory, University of California,
Berkeley, CA 94720-7450; korpela@ssl.berkeley.edu}
\altaffiltext{2}{Korea Advanced Institute of Science and Technology,
305-701, Daejeon, Korea}
\altaffiltext{3}{Korea Astronomy and Space Science Institute, 305-348,
Daejeon, Korea}

\newcommand{\iNeVI}{$\lt4100$}
\newcommand{\eNeVI}{--}
\newcommand{\mNeVI}{--}
\newcommand{\iHeII}{1850}
\newcommand{\eHeII}{$\pm$180}
\newcommand{\mHeII}{$^{+0}_{-500}$}
\newcommand{\iNIV}{1984}
\newcommand{\eNIV}{$\pm$170}
\newcommand{\mNIV}{$\pm$450}
\newcommand{\iNIII}{$\lt 2300$}
\newcommand{\eNIII}{--}
\newcommand{\mNIII}{--}
\newcommand{\iNII}{8500}
\newcommand{\eNII}{$\pm$730}
\newcommand{\mNII}{--}
\newcommand{\iAlII}{5608}
\newcommand{\eAlII}{$\pm$525}
\newcommand{\mAlII}{--}
\newcommand{\iCIII}{1260}
\newcommand{\eCIII}{$\pm$270}
\newcommand{\mCIII}{--}
\newcommand{\iCIV}{5820}
\newcommand{\eCIV}{$\pm$280}
\newcommand{\mCIV}{$^{+0}_{-1490}$}
\newcommand{\iOVI}{5724}
\newcommand{\eOVI}{$\pm$570}
\newcommand{\mOVI}{$\pm$1100}
\newcommand{\iOIII}{$\lt 500$}
\newcommand{\eOIII}{--}
\newcommand{\mOIII}{--}
\newcommand{\iOIV}{1980}
\newcommand{\eOIV}{$\pm$220}
\newcommand{\mOIV}{$\pm$300}
\newcommand{\iSiII}{2430}
\newcommand{\eSiII}{$\pm$200}
\newcommand{\mSiII}{$^{+630}_{-0}$}
\newcommand{\iSiIV}{1430}
\newcommand{\eSiIV}{$\pm$120}
\newcommand{\mSiIV}{$\pm$160}

\begin{abstract}
We present \SPEARFIMS\ far-ultraviolet observations near the North Ecliptic Pole.
This area, at $b \sim$ 30\degrees\ and with
intermediate \HI\ column,
seems to be a fairly typical line of sight that is
representative of general processes in the diffuse ISM.  
We detect a surprising number of 
emission lines of 
many elements at various ionization states representing gas phases from the warm neutral medium (WNM) to the hot ionized
medium (HIM). We also detect fluorescence bands of \Htwo, which may be due to the
ubiquitous diffuse \Htwo\ previously observed in
absorption.
\end{abstract}


\section{Introduction}

The North Ecliptic Pole (NEP) is a
region of the sky with no obviously unusual features.  
Many prior space-missions (\Einstein, \ROSAT, \COBE, 
\IRAS, etc.) have conducted deep surveys of this region.
%, most often in search of extragalactic point sources. 
Several ground based surveys have also
concentrated on this region as a complement to the space-based
surveys \citep{labov89,elvis94}.
Its location at
moderate galactic latitude (b=+29\degrees) places it clear of the bright stars
of the galactic plane.  
%Its galactic longitude (l=96\degrees) allows
%velocities due to galactic rotation to be used to separate interstellar 
%features according to distance.  
In this region, the Galactic N(H{\sc I}) 
varies from about 2\tten{20}\persqrcm\ to 8\tten{20}\persqrcm which
corresponds to an dust opacity of \taudust=0.6-2.3 at
1032\AA \citep{sasseen02} or \taudust=0.3-1.4 at 1550\AA
\citep{savage79}.

The Spectroscopy of Plasma Evolution from Astrophysical Radiation 
(\SPEAR) instrument, 
designed for observing emission lines from the diffuse ISM 
was launched in late 2003.
%on the STSAT-1 satellite,
\SPEAR\, is a dual-channel FUV imaging spectrograph
(short-$\lambda$ (S) channel 900 - 1150 \AA, 
long-$\lambda$ (L) channel 1350 - 1750 \AA)
with 
$\lambda/\Delta\lambda\sim$550,
with a large field of view 
(S: 4.0$^{\circ} \times$ 4.6', 
L: 7.5$^{\circ} \times$ 4.3')
imaged at 10$\arcmin$ resolution.
See \cite{edelstein05a, edelstein05b} for an overview of the
instrument, mission, and data analyses.
%The wide field of view and
%$\sim10\arcmin $ spatial resolution of the \SPEAR\ instrument
%make it uniquely capable of studying the
%emission from the diffuse interstellar medium (ISM). 
\SPEAR\ sky-survey observations consist of sweeps at constant ecliptic
longitude from the
NEP to the south ecliptic pole through the 
anti-solar point.  The duration of each sweep varies between 900 and 1500
s. 
%One result of this scanning method is that
Each survey scan 
includes the region near the north and south ecliptic poles resulting in 
large exposure times in these regions.  
We report on deep \SPEAR\ observations of FUV emission spectra from a
15\degree\ radius region centered on
the NEP.

\section{Observations and Data Reduction}

The NEP data analyzed here include sky-survey and pointed observations that
occurred between 8 November 2003 to 10 November 2004.
To remove times of high background and bright stars from the data set, we have
excluded times during which the count rate exceeds 100
s$^{-1}$. (In comparison, a typical photon count rate is $\simlt 20$  s$^{-1}$.)
Because stars transit the instrument slit in $\sim 5$ s during
survey sweeps, this results in minimal loss of observing time.  
The area of sky lost due to a star is similar to  the instrument
imaging resolution across the slit width ($\sim 10 \arcmin \times 5
\arcmin$). See \cite{edelstein05a} for a description of the data reduction
pipeline.

To further remove stellar contamination from our dataset, 
we excluded locally intense spatial bins ($\sim$30\% of the viewed area), 
defined to have $\simgt3\times$ the median count rate.
This level corresponds with a slope change in the 
histogram of the log count rate,
indicative of a difference in distribution of scattered vs direct starlight. 
Inclusion of unresolved starlight should not
be a factor other than to raise emission-line determination errors.
To obtain the highest sensitivity possible, we have binned the entire dataset
into a single spectrum including a total of 3.5\tten{5} counts 
over an effective full-slit exposure of 13.5 ks. 

\section{Spectral Modeling}

\newcommand{\figureone}{
\ifpp{
\begin{figure*}[tb]
}
\ifms{
\begin{figure}[p]
}
\plotone{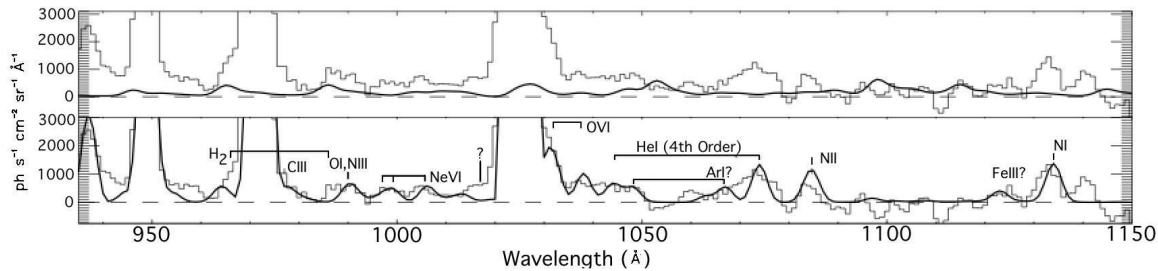}
\caption{
S channel \Htwo\ and line emission models.
Upper panel: best fit \Htwo\ fluorescence model is shown against the residuals
following subtraction of the continuum/background model.  The \Htwo\  
features' positions are well predicted, but their ratios are not. 
The \Htwo is oversubtracted near 1100 \AA\ and undersubtracted 
shortward of \Lyb \singlet 1026.
Features at $\lambda <$ 1000 \AA\ are 
underpredicted. Lower Panel: The best fit atomic line spectrum is plotted against the residuals
following subtraction of the \Htwo spectrum.  
Probable \Htwo features  and several as yet unidentified features 
remain throughout the band.\label{short}
}
\ifpp{
\end{figure*}
}
\ifms{
\end{figure}
}
}

\ifpp{
\figureone
}

The raw \SPEAR\ spectrum 
%shown in Figure~\ref{spectrum}
is composed of many components:~detector background, 
dust-scattered stellar continua, direct and instrumentally
scattered airglow, and IS atomic and molecular emission.
%in which we are most interested.  
 To model a spectrum we simultaneously fit all 
of the components to the observed spectrum.  

%In addition to the components, 
The spectral shape is distorted by the detector electronics
and by the data processing pipeline. 
These distortions appear as fluctuations in 
a flat-field image.
Any broad-band component used in our model spectrum must
be multiplied by this nonlinearity function.
\citep{korpela03, rhee02}. 

Both S and L are affected 
by instrumental background
due to cosmic rays, radioactive decay within the detector, and 
thermal charged particles entering the instrument.  These sources of 
background are relatively uniform across the face of the detectors.  Although
the charged particle rate can change with time and orbital position, the
positional distribution of these background events is dominated by static
fields within the instrument and is relatively constant.  
A sum of many
shutter-closed dark exposures multiplied by a rate 
factor can be used to model this
contribution to the spectrum.

In L, the largest broad-band spectral component is
dust scattered stellar continua.  We model the input spectrum to the dust
scattering process as a sum of 
spectra for upper main sequence (UMS) stars.  Each spectrum is weighted by using
a power law UMS luminosity function 
(${dN}/{dM_v}\propto 10^{\alpha M_v}$, where $M_v$ is
the absolute magnitude and 
${dN}/{dM_v}$ is the volume density of stars per
unit magnitude)
and a dilution
factor dependent upon the weighted mean distance of the stars.
The spectral sum is multiplied by a dust opacity
scaled to a variable $N(HI)$ and by a dust
albedo function \citep{draine03} with 
a variable total dust surface area.  The best fits for this 
function typically occur very near to the measured UMS 
luminosity function with $\alpha \sim 0.2$ \citep{reed01}.  
Because the stellar features in this spectrum cannot 
be directly measured, there may be 
some bias in our measurements of the overlying
resonance lines (\CIV\ \doublet\ 1548,1551, \OVI\ \doublet 1032,1038, \SiIV
\doublet 1394,1403, etc.) of order $\sim 1000$ \lu (LU).  We
include this effect in our error analysis.  Our best fit continuum,
which could include unresolved stars,
is $300\pm^{140}_{50}$ \cu\ (CU) in L. In S, due to its lower
sensitivity and higher non-astrophysical background, we are only able to place
a statistical upper limit of $1200$ CU. However, given the wavelength
dependence of the dust albedo, we expect that the true S 
astrophysical continuum is no higher than in L.

We first attempted to fit atomic line emission
by using linear combinations
of collisional ionization equilibrium (CIE) plasma models with a limited
abundance gradient vs $T$.
We attribute this method's lack of success 
to photoexcitation of resonance lines, 
non-CIE effects and abundance variations that are not $T$ related.
To fit the line emission we have, instead, fit the spectrum of each species
separately with a per-species intensity and $T$ parameter.  The
emission line spectra are convolved with a $\lambda$-dependent Gaussian 
to model the instrument spectral resolution.
We use the atomic parameters of CHIANTI \citep{young03} to determine the
spectrum of the species vs $T$.  This spectrum is then normalized
and multiplied by the intensity parameter.  When CHIANTI parameters
are not available for a species, we use a spectrum calculated using the 
NIST Atomic Spectra Database.

We then consider the effects of self-absorption of spectral resonance lines.  
In the
case of highly ionized species we assume all lines are optically thin.   We also
assume this to be the case for species that would be completely photoionized in the
WNM (ionization potential (IP) $\lt 13.6\eV$). For
species which are likely to be abundant in the WNM and warm ionized medium 
(WIM), 
(IP $\gtsim 13.6 \eV$) we assume that resonance lines are optically thick
($\tau \gt 10$) and that the distances to the illuminating stars are large
compared to the $\tau = 1$ surface.  In cases where the 
ground state is split, absorbed resonance
line photons will be converted to the excited state, greatly
changing the relative contributions in the multiplet.  We model this by
attenuating the ground state transition by a large factor.  We
will develop a more accurate way of treating these optical depth effects
in future work.  

\section{Discussion}

%\subsection{Geocoronal Emission}

The spectrum of the bright Lyman-series geocoronal emission,
easily visible in Fig.~\ref{short},  can be 
modeled using a temperature and total line flux.  
The largest S broadband component is scattered geocoronal \Lya \singlet
1216 emission.
This line occurs in the bandpass gap between S and L.
L is immune to \Lya\ contamination because it includes a \CaF filter
which is very nearly opaque to \Lya radiation.
We fit the S scattering with two components,
an exponential scatter $\propto exp(-\left|\lambda-\lambda_0\right|/\alpha)$ 
with $\alpha \sim 200$\AA, and an isotropic component.  The scattering,
%amplitude of both of the components 
summed over the entire spectrum, 
is $\sim$ 0.08\% of a $\sim$5 kRy \Lya incident flux.   
Because the \Lya transition is optically thick, 
and because the scattering fractions are independent parameters, 
we fix the estimated incident \Lya flux at 1100$\times$ 
the \Lyb flux which helps the
solution to converge.  While the
fractional scatter and exponential width of 
these components are free parameters in our spectral model, 
the values we obtain vary little from one dataset to the next.

Our model includes spectral features of atmospheric \OI, \CI, and \NI\ 
that are much fainter than when seen in day-time observations
(e.g. \FUSE), as \SPEAR\ only observes during orbital eclipse.  
We detect a spectral feature near 990 \AA\ that is
consistent with the \OI\ \singlet 989 transition.  However, this feature 
may be blended with \NIII\ \singlet 990.  If the entirety of the 990
\AA\ feature is due to \OI, the lack of detectable \OI\ \singlet 1356
emission would indicate $T>10^{3.6}$ K, which is plausible for
interstellar \OI.   We do detect \OI\ \singlet 1356 in data that is not count
rate filtered.
The other possibility is that the 990 \AA\ is due to \OI\ at
$T\sim$ few hundred K which seems unlikely even for atmospheric
\OI.  The most likely explanation for the lack of any detection of
\OI\ emission at $\lambda \neq 990 \AA$ is that a large
portion, if not all, of the 990 \AA\ emission is due to \NIII.  

%\subsection{Molecular Fluorescence}

\newcommand{\figuretwo}{
\ifpp{
\begin{figure*}[tb]
}
\ifms{
\begin{figure}[p]
}
\plotone{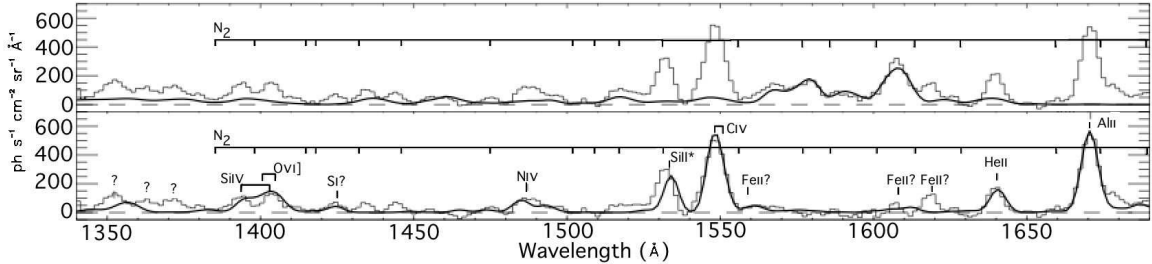}
\caption{
L channel \Htwo\ and line emission models.
\label{long}
}
\ifpp{
\end{figure*}
}
\ifms{
\end{figure}
}
}
\ifpp{
\figuretwo
}

We detect relatively strong
emission from \Htwo\ fluorescence bands even in this region
of intermediate $N(HI)$.   This is not entirely surprising since absorption
due to \Htwo\ is a ubiquitous feature of FUV spectra taken by \FUSE.
\citep{shull00} We fit these
features by using interpolation between of a sparse grid of \Htwo\ models
(B. Draine 2004, private communication).
The upper panels of Figs.~\ref{short} \& \ref{long} show the best fit \Htwo\ model plotted against the residuals of
the spectrum following background, continuum, and scattering subtraction.
We detect H$_{2}$ emission features in both spectral bands.  In 
S we
detect the \Htwo\ \doublet 965 band (to the left of the HI \singlet 972
line) and the \Htwo \doublet 986 band (to the left of the \OI,\NIII\ \doublet 989
feature). The \Htwo \doublet 1053 and \doublet 1097 bands are not detected at any
significance.
In L, H$_{2}$ provides 
contributions throughout the spectrum.  We attribute the triple
peaked band structure between 1350 and 1380 \AA, the peak near 1440
\AA, the broad features \doublet 1520-1530, \doublet 1570-1590, and the
peak near 1608\AA\ to H$_{2}$.  Note that these features are not well fit
by our \Htwo\ model.  It is not surprising that our sparse
linear model with a single dust absorption column does not fit these data
since the observed emission derives from
many lines of sight with varying illumination, gas
temperatures and absorbing columns. 
In Figs.~\ref{short} \& \ref{long},  lower panels  we show the residual spectrum following subtraction of the
H$_{2}$ component. 

We see intriguing indications of \Ntwo fluorescent emission 
, most notably the features at 1434\AA, 1445\AA,
1475\AA, and 1660\AA.  \FUSE\ has detected absorption due to \Ntwo
along some sight-lines, albeit
at shorter wavelengths \citep{knauth04}. We cannot yet discount the
possibility that these \Ntwo features are atmospheric, as near the magnetic poles,
\Ntwoplus can be transported to high altitude.  Further study to
determine whether the \Ntwo emission varies with spacecraft orbital position 
or with galactic latitude of the observation should resolve this question.
If confirmed as astrophysical, this would be the first detection of fluorescent
\Ntwo emission from the diffuse ISM.

%\subsection{Emission from the Warm Neutral and Warm Ionized Media}

\SPEAR\ has discovered some surprising spectral features, such as the resonance
lines of the singly ionized species, \SiII\ and \AlII.  The \AlII\ feature at 1671
\AA\ very nearly overlies the \OIIIb\ \doublet 1665 which we do not
detect in this dataset.  In a lower resolution instrument these spectral
features could be blended resulting in confusion of the \OIIIb\ and \AlII\ 
emission.
We think it is likely that at least some of the prior claimed detections
\citep{mb} of 
\OIIIb\ were in fact misidentified detections of \AlII\ \doublet 1671.  
We attribute the feature at 1533 \AA\ to the \SiII$^{*}$ \singlet 1533.4 transition into
the upper fine structure level of the ground state.  The offset between the
model feature and the observed line may be due to an uncorrected continuum
feature underlying this line (Fig.~3 of \cite{edelstein05a}).
Since \SiII\ is the predominant state of silicon in
both the WNM and WIM, the true \SiII\ \singlet 1526.7
ground state transition is likely to be optically thick and therefore
undetectable.  
Our spectral models are improved by the addition
of \FeII\ transitions throughout band, most notably 
the \FeII\ \doublet 1564, which
fills the gap between \CIV\ \doublet 1548,1551 and the \doublet 1570-1590
H$_{2}$ band.  It is also possible that there is a contribution to
the \Htwo \doublet 1600-1620 band due to the \FeII\ \doublet 1608 resonance lines.
These lines have a fairly low oscillator strength and are expected to be
optically thin along this line of sight.  However, given the low oscillator
strength, we expect these lines to be faint compared to the \SiII\
and \AlII\ features.

\newcommand{\tableone}{
\begin{deluxetable}{lrrcc}
\tablecaption{Observed warm/hot ISM emission line intensities
\label{him_lines}}
\tablehead{
\colhead{Species} & \colhead{$\lambda$} & \colhead{$I_\lambda$} & 
\colhead{Statistical} & \colhead{Model} \\
\colhead{\ } & \colhead{\ } & \colhead{\ } & \colhead{Error} &
\colhead{Uncertainty} \\
\colhead{\ } & \colhead{\AA} & \colhead{LU$^a$} & \colhead{LU} & \colhead{LU} 
}
\startdata
\OIIIb & 1667 & \iOIII$^b$ & \eOIII & \mOIII \\
\OIVb & 1400 & \iOIV & \eOIV &\mOIV \\
\OVI & 1032,1038 & \iOVI & \eOVI &\mOVI \\
\CIII & 977 & \iCIII & \eCIII & \mCIII \\
\CIV & 1548,1551  & \iCIV & \eCIV & \mCIV \\
\SiII$^{*}$ & 1532 & \iSiII & \eSiII & \mSiII \\
\SiIV & 1394,1403 & \iSiIV & \eSiIV & \mSiIV \\
\NII & 1085 & \iNII & \eNII & \mNII \\
\NIII & 990 & \iNIII & \eNIII & \mNIII \\
\NIV & 1485 & \iNIV & \eNIV & \mNIV \\
\AlII  & 1671 & \iAlII & \eAlII & \mAlII \\
\HeII & 1640 & \iHeII & \eHeII & \mHeII \\
\NeVI & 999,1005 & \iNeVI & \eNeVI & \mNeVI \\
\enddata
\tablecomments{$^a$LU=\lu\ \ $^b$Upper limits are 90\% confidence}
\end{deluxetable}
}

\ifpp{
\tableone
}

%\subsection{Emission from the ``Hot'' Ionized Media}
Emission lines of highly ionized gas expected to be generated
in a CIE plasma at HIM temperatures
are shown in Table \ref{him_lines}.
Of these we  detect ($> 3\sigma$) \CIII\ \singlet 
977, \CIV\ \doublet 1548,1551, \SiIV\ \doublet 1394,1403 and
\OVI\ \doublet 1032,1038.  
The \CIV\ \doublet 1549 and \SiIV\ \doublet 1400 emission
overlie prominent stellar absorption features which are likely to be present in
the dust-scattered stellar radiation. 
Since the depth of these features is
unknown we set the lower range of our model uncertainty to be the intensity that
would be calculated if there were 
no continuum feature present.  

Many of these emission features are resonance lines (R.~J. Reynolds 2005,
private communication)
and, because of the substantial IS radiation field at these 
wavelengths, further work will be required to determine whether 
these lines arise due to collisional excitation
in the HIM or due to resonant scattering of the IS
radiation field.  We believe that the short wavelength lines are less likely to
be resonant scatter because the relative intensity of the scattered stellar 
continuum in the S band is much lower \citep{korpela98}.

Our \CIV\ \doublet 1548,1551 intensity of \iCIV\eCIV\ LU is
marginally higher than values for the
diffuse ISM reported by \cite{mb}.  However Martin and Bowyer assumed a
featureless continuum rather than a continuum with stellar features.  
If we make the same assumption our value drops to
4330$\pm$210 LU~which is more within the range of values they report.

Our \OVI\ \doublet 1032,1038 intensity of \iOVI\mOVI\ LU~is somewhat 
higher than the typical value measured by \FUSE\  \citep{shelton}.  This
could be a systematic effect due to the uncertain calibration of the
S band. In fact the ratio of the \Htwo\
feature intensities between L and
S\ suggests that we have underestimated
the sensitivity of S.
Since this is an average over a very large
region, it is also possible that the \OVI\ emission is due to several small
regions with very high intensity, such as SNR.  

The double peaked feature near 1400 \AA\ is a blend of \SiIV\ \doublet\
1393.8,1402.8 and \OIVb\ \doublet\ 1400.7,1407.4.  The \OIVb\ emission is of
great interest because, as a semiforbidden line, it is not strongly affected by
the radiation field, but is a direct indicator of collisional processes.
Assuming
a 2:1 ratio for the \SiIV\ doublet components, our model determines
the \OIVb\ doublet
intensity to be \iOIV\mOIV\ LU~and the \SiIV\ doublet intensity to be
\iSiIV\mSiIV\ LU.

We detect high stage ions of two noble gases in our spectra, \HeII\ \doublet\
1640 and \NeVI\ \doublet\ 999,1005.   Although the features corresponding to the
\NeVI\ lines are statistically significant, the poor fit to the underlying 
\Htwo\ bands prevents us from placing more than an upper limit to these
features.
Because of the lack of detection of any \OI\ \singlet\ 1356,
we do not expect any contamination of the \HeII\ line with \OI\ \singlet\ 1641.
The \HeII\ line is much brighter than would
be expected in a solar abundance CIE model (in ratio to \CIII\ and \OIII).
This could be due to depletion of C and O to $\simlt 0.1$ solar.
A depletion of this magnitude would be significantly higher than the accepted
values for the ISM.

\newcommand{\figurethree}{
\begin{figure}[tb]
\plotone{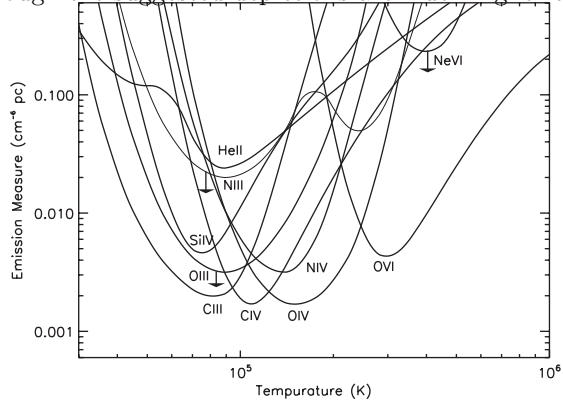}
\caption{Solar abundance emission measure (EM) vs $T$ for detected emission lines and
significant upper limits.  The lines represent the loci of EM 
necessary to produce the observed intensities.\label{EM}}
\end{figure}
}
\ifpp{
\figurethree
}
In Fig.~\ref{EM} we show equivalent solar abundance emission measure (EM) and limits
for several species.  The results are not entirely 
consistent between species.  For example higher EM is necessary
to explain the \SiIV\ and \HeII\ emission than are compatible with the amount of \CIII\
emission. The questionable \NeVI\ detections would indicate EM 
10$\times$ that suggested by the \OVI\ measurements.  The bulk of
the inconsistency in \NeVI\ is likely to be due to undersubtraction of the 
underlying H$_{2}$.
Much of the inconsistency in other lines is likely due to
non-CIE effects which are present in gas being heated or cooling
at these temperatures.  
Some portion of the difference could be due to abundance variations (with ISM depletions of O and
C higher than those of Ne, He, and Si), although the suggested depletions are
much higher than the accepted ISM values.  
In future work we will
derive constraints to non-CIE cooling models and abundances using these line
ratios.

\section{Conclusions}

We have presented the FUV spectra of a 30\degree\ region around the 
NEP.  We detect a variety of emission lines from high and low
ionization gas.  The high ionization lines, taken on a per species basis,
are consistent with 
EM from 0.001 to 0.005 \emunits\ throughout 
$T=10^{4.5}$ to 10$^{5.5}$ K range.   For those few species which have multiple
lines in the band, the calculated temperatures tend to fall near the CIE peak
abundance for the species.  Despite this, a linear combination of
CIE models with a restricted abundance vs $T$ gradient 
does not provide a good fit to the spectrum.  This is
likely due to non-CIE effects and abundance variations are not
directly $T$ related.  It is also likely that many lines include some 
resonantly scattered stellar radiation.  Modeling this scattering 
will be a significant portion of our continued research.

We have discerned line emission from species that are expected
to exist in the
WNM and WIM.  We interpret these lines as resonantly
scattered stellar radiation.  In the future, we will investigate if spatial 
variation of the \AlII\ and \SiII\ features can indicate the effect of
grain formation and destruction on the gas phase abundances of these elements.

Despite relatively low galactic $N(HI)$ in this direction, we see a
substantial amount of \Htwo\ fluorescence.  Further modeling of the \Htwo\ 
fluorescence is necessary to fully understand faint spectral features
which could be related to \Htwo\ or due to unrelated atomic lines. 

\acknowledgements
\SPEARFIMS\ 
is a joint project of KASSI \& KAIST (Korea) and U.C., Berkeley (USA),
funded by the Korea MOST and NASA Grant NAG5-5355.

\ifms{
%\newpage
%\figureone
\newpage
\figureone
\newpage
\figuretwo
\newpage
\figurethree
\newpage
\tableone
}

\end{document}